\documentclass[12pt,preprint]{aastex}

\shorttitle{Progenitor of SN 2006my}\shortauthors{Leonard et al.}

\begin{document}

\title{An Upper Mass Limit on a Red Supergiant Progenitor for the Type
  II-Plateau Supernova SN 2006my\footnote{Some of the data presented herein
  were obtained at the W.M. Keck Observatory, which is operated as a scientific
  partnership among the California Institute of Technology, the University of
  California and the National Aeronautics and Space Administration. The
  Observatory was made possible by the generous financial support of the
  W.M. Keck Foundation.  Additional data were obtained from the data archive at
  the Space Telescope Science Institute. STScI is operated by the Association
  of Universities for Research in Astronomy, Inc., under NASA contract NAS
  5-26555. }}


\author{Douglas C. Leonard\altaffilmark{1},
Avishay Gal-Yam\altaffilmark{2},
Derek B. Fox\altaffilmark{3},
P. B. Cameron\altaffilmark{4},
Erik M. Johansson\altaffilmark{5},
Adam L. Kraus\altaffilmark{4},
David Le Mignant\altaffilmark{5},
Marcos A. van Dam\altaffilmark{5} }

\altaffiltext{1}{Department of Astronomy, San Diego State University, San
  Diego, CA 92182-1221; leonard@sciences.sdsu.edu}

\altaffiltext{2}{Benoziyo Center for Astrophysics, Weizmann Institute of
  Science, 76100 Rehovot, Israel} 

\altaffiltext{3}{Department of Astronomy and Astrophysics, Pennsylvania
State University, 525 Davey Laboratory, University Park, PA 16802}

\altaffiltext{4}{Department of Astronomy, MS 105-24, California Institute of
  Technology, Pasadena, CA 91125}

\altaffiltext{5}{W. M. Keck Observatory, 65-1120 Mamalahoa Highway, Kamuela, HI
  96743}


\begin{abstract}

We analyze two pre-supernova (SN) and three post-SN high-resolution images of
the site of the Type II-Plateau supernova SN~2006my in an effort to either
detect the progenitor star or to constrain its properties.  Following image
registration, we find that an isolated stellar object is not detected at the
location of SN~2006my in either of the two pre-SN images.  In the first, an
$I$-band image obtained with the Wide-Field and Planetary Camera 2 on board the
{\it Hubble Space Telescope}, the offset between the SN~2006my location and a
detected source (``Source 1'') is too large: $\geq 0.08^{\prime\prime}$, which
corresponds to a confidence level of non-association of $96\%$ from our most
liberal estimates of the transformation and measurement uncertainties.  In the
second, a similarly obtained $V$-band image, a source is detected (``Source
2'') that has overlap with the SN~2006my location but is definitively an 
extended object.  Through artificial star tests carried out on the precise
location of SN~2006my in the images, we derive a $3\ \sigma$ upper bound on the
luminosity of a red supergiant that could have remained undetected in our
pre-SN images of $\log L/L_\odot = 5.10$, which translates to an upper bound on
such a star's initial mass of $15 {\rm\ M}_\odot$ from the STARS stellar
evolutionary models.  Although considered unlikely, we can not rule out the
possibility that {\it part} of the light comprising Source 1, which exhibits a
slight extension relative to other point sources in the image, or {\it part} of
the light contributing to the extended Source 2, may be due to the progenitor
of SN~2006my.  Only additional, high-resolution observations of the site taken
after SN~2006my has faded beyond detection can confirm or reject these
possibilities.

\end{abstract}


\section{Introduction}
\label{sec:1}

The most common class of core-collapse supernovae (SNe) displays a distinct
plateau in its optical light curve, and is therefore dubbed Type II-Plateau
(II-P; see \citealt{Filippenko97} for a review of SN classifications).  This
type of stellar explosion has long been thought to result from the
core collapse and subsequent envelope ejection of isolated red supergiant (RSG)
stars, but it is only in recent years that direct observational evidence of the
progenitor-SN~II-P connection has begun to accumulate
\citep{Vandyk03,Smartt04,Maund05,Li06a,Eldridge07,Schawinski08}.

By registering pre-SN and post-SN images, usually taken at high resolution
using either space-based optical detectors, or ground-based infrared detectors
equipped with laser guide star adaptive optics systems, progenitor star
identifications have now been proposed for seven SNe~II-P (for a contemporary
review, see \citealt{Smartt08}, and references therein).  Although different in
detail, all seven proposed progenitors have properties consistent with those of
supergiant stars.  Because the field is still in its infancy --- at this point
none of the proposed progenitor objects has been definitively confirmed by
its absence in high-resolution images of the SN site taken after the SN has
faded beyond detection --- it is imperative to carefully examine every new
progenitor claim.  Here we investigate the status of the progenitor of
SN~2006my, for which \citet[][hereafter L07]{Li07} recently proposed the
identification of an RSG progenitor in pre-SN images and derived a zero-age
main-sequence mass for it of $M_{\rm ZAMS} = 10^{+5}_{-3} {\rm\ M}_\odot$.

Although discovered several months after explosion \citep{Stanishev06}, the
photometry and spectroscopy of SN~2006my presented by L07 clearly establish it
as an SN~II-P.  To identify the progenitor star, L07 register post-SN
ground-based optical images taken with the Canada-France-Hawaii Telescope in
the Sloan $r^\prime$ band under excellent seeing conditions (typical stellar
full width at half-maximum [FWHM] $= 0.6^{\prime\prime}$) with pre-SN {\it
Hubble Space Telescope} ({\it HST}) Wide-Field and Planetary Camera 2 (WFPC2)
images (typical stellar FWHM $= 0.15^{\prime\prime}$ on the WF2 chip, on which
the SN site is located), and identify a source in the pre-SN images within the
$1\ \sigma$ error circle estimated from the transformation uncertainty.  Here, we
reexamine this identification with the benefit of two new sets of
high-resolution, post-SN images: one taken with the wide-field channel of the
Near Infrared Camera 2 operated behind the laser guide star assisted adaptive
optics system (\citealt{Wizinowich06}) on the Keck II 10-m telescope (stellar
FWHM $= 0.10^{\prime\prime}$; note that SN~2006my served as its own tip-tilt
star for the observations), and the other taken with the {\it HST} WFPC2 camera
(with the SN centered on the PC chip; typical stellar FWHM $=
0.08^{\prime\prime}$) as part of a study of the progenitors of core-collapse
supernovae (GO 10803; PI: Smartt).

This paper is organized as follows.  In \S~\ref{sec:2.1} we present the pre-SN
and post-SN images and details of the photometric measurements performed on
them; in \S~\ref{sec:2.2} we describe the image registration process along with
our conclusion that no isolated stellar source is detected at the location of
SN~2006my in either of the pre-SN images; in \S~\ref{sec:2.3} we estimate
detection limits in the pre-SN images; and in \S~\ref{sec:2.4} we derive an
upper mass limit on the progenitor of SN~2006my under the assumption that it
was a RSG.  We summarize our findings in \S~\ref{sec:3}.

\section{Data Analysis}
\label{sec:2}

\subsection{Pre-SN Image Photometry with HSTphot}
\label{sec:2.1}

Table~\ref{tab:1} lists information on the five sets of image data considered
by our study, hereafter referred to by the designation assigned to the final
combined image from each data set.  We first discuss our photometry of the two
pre-SN images, V1 and I1.  The individual frames comprising these datasets were
preprocessed through the standard Space Telescope Science Institute pipeline
using the latest calibrations as of 2008 July 23. The images were then further
processed following the procedure described by \citet{Leonard10}, which employs
the suite of programs designed specifically for the reduction of WFPC2 data
that are available as part of the HSTphot \citep{Dolphin00a} software package
(ver. 1.1.7b; our implementation includes the most recent update of 2008 July
19).  The {\it hstphot} task automatically accounts for WFPC2 point-spread
function (PSF) variations, charge-transfer effects across the chips,
zeropoints, and aperture corrections.

We performed photometry on the final, combined images using the {\it hstphot}
task with option flag 10, which combines turning on local sky determination,
turning off empirically determined aperture corrections (using default values
instead), and turning on PSF residual determination. We ran {\it hstphot} on V1
and I1 individually with a signal-to-noise ratio (S/N) threshold of 1.0.

In addition to flight-system magnitudes and uncertainties, {\it hstphot}
returns several measurement parameters for each object detected.  For our
purposes, the most important of these are the ``object type'' and ``sharpness''
parameters.  To determine the object type, {\it hstphot} compares the
goodness-of-fit (i.e., the $\chi$ value; see \citealt{Dolphin00a}) of the
detected source's spatial profile with three different models: (1) the best
stellar profile (i.e., the library PSF + residuals); (2) a single ``hot'' pixel
without background; and (3) a completely flat profile.  If the stellar profile
provides the best match, the object is labeled a ``star'' and designated as
object Type `1', `2', or `3', depending on whether it is deemed a ``good
star'', a ``possible unresolved binary'', or a ``bad star' (a star centered on
a bad pixel), respectively.  If the single ``hot'' pixel provides the best
profile match, the object is labeled a ``single-pixel cosmic ray or hot pixel''
and designated as object Type `4'.  Finally, if a flat profile matches best,
the object is labeled an ``extended object'' and designated as object Type `5'.
\citet{Dolphin00a} notes that because this is, by design, a conservative test for
object type discrimination (i.e., it is a high threshold to be classified as
anything other than a good star), many nonstellar objects will still be
classified as Type 1 objects. For particular objects of interest, therefore,
examination of the ``sharpness'' parameter is recommended, where sharpness
values between -0.3 (object PSF broader than library PSF) and +0.3 (object PSF
narrower than library PSF) indicate confident point-source detections
\citep{Dolphin00a}.

\begin{figure}
\scalebox{1.1}{
\plotone{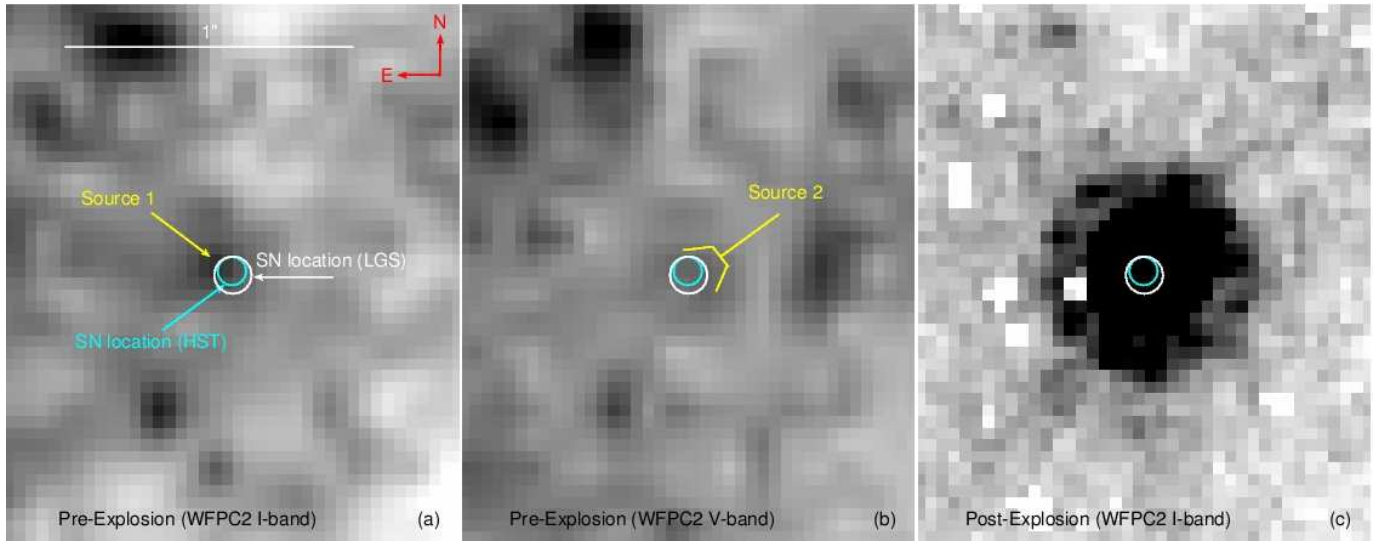} 
              }
\caption{Site of SN~2006my in pre-SN images I1 (panel {\it a}; all image
  designations are as given in Table~\ref{tab:1}) and V1 (panel {\it b}), and
  post-SN image I2 (panel {\it c}).  All images have been transformed and
  resampled to the pixel grid of the K1 image (\S~\ref{sec:2.2}).  The cyan
  circles indicate the approximate $5\sigma$ uncertainty (for a two-dimensional
  Gaussian; see the Appendix) of the position of SN~2006my relative to the
  transformed I1 image as measured in the transformed I2 image.  The white
  circles indicate the same level of uncertainty in the location of SN~2006my
  as measured in the K1 image relative to the transformed I1 image.  Two
  ``sources of interest,'' discussed in the text, are labeled Source 1 and
  Source 2 in panels {\it a} and {\it b}, respectively.
\label{fig:1} }
\end{figure}

Figures~\ref{fig:1}({\it a}) and \ref{fig:1}({\it b}) show an $\sim
1^{\prime\prime}$ region surrounding the site of SN 2006my on pre-SN images I1
and V1 (respectively), following transformation of both images to the pixel
grid of image K1 (see \S~\ref{sec:2.2}). The SN 2006my site is rather complex,
and while common objects are evident in Figures~\ref{fig:1}({\it a}) and
\ref{fig:1}({\it b}), the immediate vicinity of SN~2006my appears quite
different in them.  In I1, an object of Type 1 (``good star'') is reported by
{\it hstphot} close to the SN location with a S/N of 5.6; we label this object
``Source 1'' in Figure~\ref{fig:1}a.  In V1, an object of Type 5 (``extended
object'') is detected by {\it hstphot} with a S/N of 4.70; this extended source
appears to occupy a ``horseshoe-shaped'' region that includes the SN location.
We label this object ``Source 2'' in Figure~\ref{fig:1}b.  These two sources
appear to be distinct from one another in the combined images.

Source 1 is the object identified by L07 as the probable progenitor of
SN~2006my.  We confirm both the pixel location and photometry reported by L07
(also determined using HSTphot), with values measured by us (L07) of [x,y] =
[410.23, 158.59] ([410.22, 158.63]),\footnote{Note that all pixel coordinates
given in this paper are in the ``IRAF'' system --- the coordinate system
reported by using the {\it imexamine} task within IRAF or by running DAOPHOT
\citep{Stetson87} --- in which an integer value is assigned to a star that is
centered on the center of a pixel.  These pixel values are 0.5 greater in both
{\it x} and {\it y} than those reported by {\it hstphot}, which follows the
DoPHOT \citep{Schechter93} convention of assigning an integer value to a star
that is centered in the lower left corner of a pixel.} and flight-system
magnitude F814W = $24.48 \pm 0.19$ mag ($24.47 \pm 0.20$ mag), which
corresponds to $I = 24.46 \pm 0.19$ mag ($24.45 \pm 0.20$ mag) following
transformation according to the prescriptions of \citet{Holtzman95a} and
\citet{Dolphin00b}.  As noted by L07, although formally classified as an object
of Type 1 (``good star'') by {\it hstphot}, Source 1 exhibits a slight
east-west extension in the original WFPC2 image (see Fig.~\ref{fig:1}({\it a});
see also Figure 6 of L07). While L07 conclude that this object is most likely a
single star, we note that {\it hstphot} reports a ``sharpness'' value of
$-0.35$ for it, which places it beyond the limits suggested by
\citet{Dolphin00a} for confident point-source detections.  Source 2 is
unequivocally an extended source, as {\it hstphot} flags it as an object of
Type 5 and measures its profile to have a sharpness of $-0.58$, well beyond the
range for isolated star-like sources and for which {\it hstphot}'s PSF
star-fitting routines provide reliable photometry \citep{Dolphin02a}.  We note
that this object may be the ``extended source'' mentioned by L07 but not
further investigated due to its location near the $\sim 2\ \sigma$ error radius
of their derived SN pixel coordinates.  We shall return to further discuss both
of these ``sources of interest'' following a description of the image
registration process.

\subsection{Image Registration}
\label{sec:2.2}

To determine whether the progenitor star that exploded as SN~2006my is detected
in the pre-SN images, we first registered image I1 to image K1 by using the
IRAF tasks {\it geomap} and {\it geotran}, closely following the technique
described by \citet{Galyam05}.  To carry out the transformation, we used 16
common sources and obtained a final solution with a rms residual of 0.30 pixel
in {\it x} and 0.27 pixel in {\it y}, in the K1 pixel grid.  We then similarly
registered images V1, V2, and I2 to the transformed I1 image (using more than a
dozen common sources in each case), which resulted in five final images all
registered to the common K1 pixel grid.

Using the transformed images, we then measured the pixel locations of Source 1
(in image I1) and SN~2006my (in images K1, I2, and V2) using the centroiding
algorithm of the {\it imexamine} task within IRAF.\footnote{IRAF is distributed
by the National Optical Astronomy Observatories, which are operated by the
Association of Universities for Research in Astronomy, Inc., under cooperative
agreement with the National Science Foundation.}  The results returned by {\it
imexamine} depend somewhat on both the centering radius and fitting function
employed.  Thus, we applied a range of values --- centering radii of from 3--10
pixels, inclusive, using both Gaussian and Moffat profiles --- and took the
average as our ``best'' value and the measured scatter around the average as
the measurement uncertainty.  Due to the extended nature of Source 2 in image
V1, it was not possible to estimate a precise pixel location for it by using
the IRAF centroiding algorithm; instead, we ran the source-finding program
SExtractor \citep{Bertin96} on the image, which reported a nonstellar source
and pixel position at the location of Source 2.  The pixel locations,
measurement uncertainties, and transformation uncertainties of all objects are
given in Table~\ref{tab:2}.

Accounting for measurement and transformation uncertainties, we determine
offsets between Source 1 in image I1 and SN~2006my in images K1, I2, and V2, of
$0.083^{\prime\prime} \pm 0.017^{\prime\prime}$, $0.080^{\prime\prime} \pm
0.016^{\prime\prime}$, and $0.086^{\prime\prime} \pm 0.016^{\prime\prime}$,
which correspond to separations significant at the $4.8\ \sigma$, $4.9\ \sigma$,
and $5.2\ \sigma$ levels, respectively, for a two-dimensional (i.e., both $x$
and $y$) Gaussian.  This indicates non-association at greater than the $99\%$
confidence level (see the Appendix for a full discussion of how significance of
source separation is determined), implying that the object detected as Source 1
is in all likelihood not the progenitor of SN~2006my.  Due to the extended
nature of Source 2, it is not possible to derive a similarly well-quantified
offset and uncertainty estimate between it and SN~2006my, but we note that from
the location reported by SExtractor, its formal separation is only
$0.024^{\prime\prime}$, and examination of Figure~\ref{fig:1} reveals that the
location of SN~2006my is indeed coincident with part of the extended region
identified as Source 2.

To serve as a check on the results obtained by transforming all images to the
K1 frame, we also performed a direct, ``HST-only'', registration between images
V2 and V1, and I2 and I1, using the {\it hstphot}-reported pixel positions
(with an additional correction for distortion using the solutions of
\citealt{Anderson03}) for all ($> 20$) objects used to determine the
transformation. Results of these registrations are given in Table~\ref{tab:3}.
To assign a measurement uncertainty on the {\it hstphot}-measured object pixel
positions, we consulted Figure~4 of \citet{Dolphin00a}, which presents
estimates of the $1\ \sigma$ astrometry error for sources detected by {\it
hstphot} as a function of count level.  For the count level of Source 1 ($\sim
45$), a total astrometry error estimate of $\sim 0.4$ WF2 pixel, or $\sim 0.29$
pixels in both $x$ and $y$, is determined, while for SN~2006my (counts $>
20,000$) position uncertainties of only 0.03 pixels in $x$ and $y$ are derived.

Accounting for measurement and transformation uncertainties, we determine the
separation between Source 1 in image I1 in the transformed location of
SN~2006my from image I2 to be $0.082^{\prime\prime} \pm 0.031^{\prime\prime}$,
which represents an offset of $2.6\ \sigma$ for a two-dimensional Gaussian,
indicating non-association at the $96\%$ confidence level (see the Appendix).
This separation is nearly identical to those determined in the K1 pixel frame,
although the significance of the offset is lower.  This is due mainly to the
larger uncertainty adopted for the pixel location of Source 1 in I1 than was
found empirically for the I1 image transformed to the K1 pixel grid.
Registration of I1 onto K1 resamples the data onto a finer grid and therefore
likely provides a better localization of Source 1, although, as noted by L07,
resampling the data has the risk of smoothing several (extended) sources into a
point source.  We also note that a direct registration of V2 to I1 (not listed
in Table~3) yields an offset between Source 1 and SN 2006my of
$0.092^{\prime\prime} \pm 0.032^{\prime\prime}$, which represents an offset of
$2.9\ \sigma$ for a two-dimensional Gaussian and indicates non-association at the
$98\%$ confidence level.  With all of these lines of evidence, we conclude that
a $96\%$ confidence level for non-association is a conservative estimate of the
significance of the offset between Source 1 and SN~2006my.

We note that in the original I1 WFPC2 image, Source 1 presents itself as
elevated flux in two pixels (i.e., pixels [411, 158] and [410, 159]; see
Figure~6 of L07), one of which contains our derived location of SN~2006my at
its far edge (Table~\ref{tab:3}).  The slight extension of Source 1 noted
earlier (\S~\ref{sec:2.1}) raises the possibility that this source may, in
fact, contain light from more than one object --- for instance, two RSG lying
in adjacent WF pixels, or perhaps a compact star cluster --- an idea considered
by L07 but deemed improbable.  This lingering possibility can only be
definitively removed from consideration by future high-resolution imaging after
the SN has faded beyond detection.  For now, we conclude that from our
astrometric measurements, an isolated stellar object is not detected at the
location of SN~2006my in either of the two pre-SN images: In image I1, the
offset between the SN~2006my location and the source detected by {\it hstphot}
is too large, and in V1 the source that overlaps with the SN~2006my site is
extended.  The relevant question that arises, then, is this: How faint a
point source at the precise location of SN~2006my would we have confidently
detected as a point source in the pre-SN images?

\subsection{Detection Limits in Pre-SN Images}
\label{sec:2.3}

To set accurate detection limits on a point source at the location of SN~2006my
in the pre-SN images, we proceeded as follows.  First, we used the {\it
showpsf}\ task within HSTphot to produce the library PSFs appropriate for point
sources at the precise pixel locations (accurate to 0.1 pixel) of SN~2006my in
the V1 and I1 images.  Using these PSFs, we then inserted artificial stars of
known flux (corresponding to $21.5 < V < 27.0$ and $20.5 < I < 26.0$) at the SN
locations in the V1 and I1 images.  Then, we ran {\it hstphot} on these images
in exactly the same manner as we did when seeking a progenitor star in the
original images.

To be considered a confident ``detection'' of a single star at the SN~2006my
location, we demanded that the object be classified by {\it hstphot} as a
``good star'' (Type 1), have a sharpness between -0.3 and +0.3, and have a
reported pixel location no more than $1\ \sigma$ away from the nominal SN
location, where the uncertainty accounts for the measurement and transformation
uncertainties listed in Table~\ref{tab:3} as well as the astrometric
uncertainty appropriate for the object from Figure~4 of \citet{Dolphin00a}.

From this analysis, we derive detection limits of $V = 24.9 \pm 0.3$ mag and $I
= 24.4 \pm 0.2$ mag for point sources at the location of SN~2006my in the V1
and I1 images, respectively.  In both cases the limiting magnitude is set by
the sharpness parameter becoming less than -0.3 (i.e., source too extended to
be confidently considered a point source).  We note that these detection limits
are significantly shallower than those derived by L07, in which detection
limits were derived by examining the magnitudes of all $3\ \sigma$ detections in
the images.  Because we are specifically interested in the point-source
detection limits at the location of the SN, we consider our (less restrictive)
limits to more accurately reflect the relevant detection threshold.

\subsection{Properties of the Progenitor Star}
\label{sec:2.4}

To convert our detection thresholds to constraints on the initial mass of an
RSG progenitor star that could have exploded as SN~2006my, we employ the
metallicity-dependent stellar models produced with the Cambridge stellar
evolution code, STARS, the descendant of the code developed originally by
\citet{Eggleton71} and updated most recently by \citet[][see also
\citealt{Smartt08}, and references therein]{Eldridge04}.\footnote{The models
were downloaded from the code's Web site, at
http://www.ast.cam.ac.uk/$\sim$stars .}  The models follow stellar evolution up
to the initiation of core neon burning, which is likely to give an accurate
indication of the pre-SN luminosity; comparisons with other contemporary model
grids (i.e., \citealt{Heger00}; \citealt{Meynet00}) show that the endpoints for
stars in the $8 \rightarrow 15 {\rm\ M}_\odot$ range differ by at most $0.2$
dex in luminosity among the codes \citep{Smartt04}.

\begin{figure}
\begin{center}
\scalebox{0.8}{
\plotone{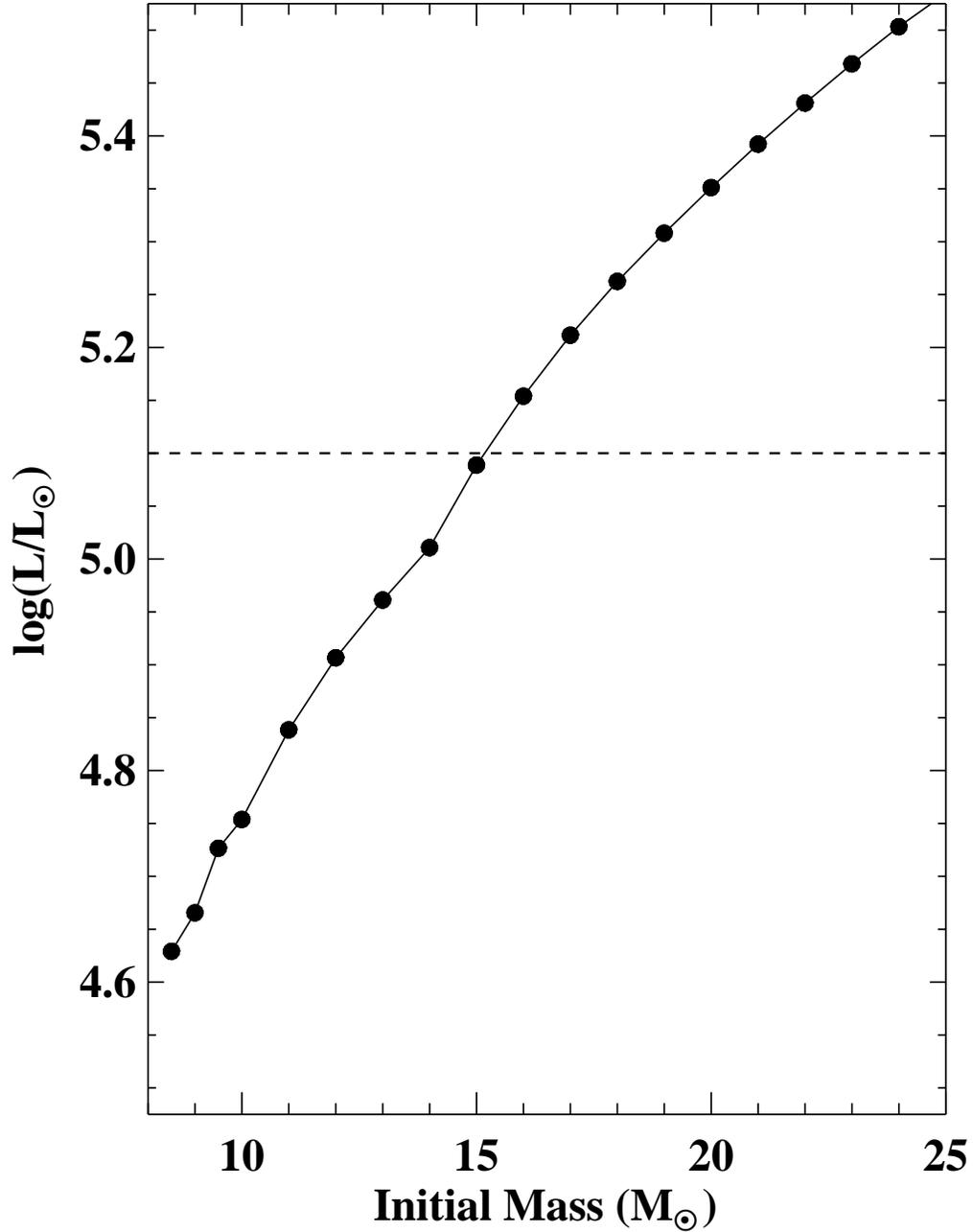} 
              }
\end{center}
\vskip 0.1in
\caption{Initial mass vs. final predicted luminosity prior to explosion for
  $Z = 0.01$ stars evolved with the STARS stellar evolution code
  \citep{Eldridge04}.  The dashed line indicates the $3\sigma$ upper luminosity
  limit for a RSG that could have remained undetected by our analysis of pre-SN
  images of the site of SN~2006my.
\label{fig:2} }
\end{figure}

Using the metallicity and radial gradient in NGC 4651 published by
\citet{Pilyugin04}, L07 derive a metallicity at the SN~2006my location of
$\log(O/H) + 12 = 8.51 \pm 0.06$, which is subsolar according to the recent
analysis of \citet{Asplund04}, who found $[O/H] = 8.66 \pm 0.05$.  We thus
use the $Z = 0.01$ (the closest metallicity calculated) STARS stellar evolution
models as the basis for deriving the upper mass limit for our study.
Figure~\ref{fig:2} displays the final predicted luminosity for stars with
masses between 8 and 20 M$_\odot$ from these models.

We now seek to determine the lowest possible luminosity that an RSG could have
had and still have been confidently detected by our analysis of the pre-SN
images.  Since it will prove to be far more restrictive, we begin by
considering the I1 image, and calculate the bolometric magnitude of RSG stars
corresponding to our detection limit through the equation:
$$M_{\rm bol} = -\mu - A_V + I + (V - I)_{\rm RSG} + {\rm BC}_V,$$
\noindent where $\mu$ is the distance modulus of NGC 4651, $A_V$ the extinction
to SN~2006my, $I$ the $I$-band detection threshold, $(V - I)_{\rm RSG}$ the
color range of RSG stars (i.e., spectral types ${\rm K3} \rightarrow {\rm
M4}$), and ${\rm BC}_V$ the bolometric correction corresponding to each $(V -
I)_{\rm RSG}$.  As detailed by L07, distance estimates to NGC 4651 span quite a
wide range.  Because it is our goal here to set the most conservative lower
bound on our detection threshold, we adopt the long distance estimate, $\mu =
31.74 \pm 0.25$ mag, found by \citet{Solanes02} by averaging seven different
Tully-Fisher distances to the galaxy.  For the extinction, we adopt $A_V = 0.08
\pm 0.02$ mag, which represents the Galactic value along the line of sight,
because there is no evidence for host-galaxy extinction (L07).  For the color
and bolometric corrections appropriate for RSG stars, we consult the values
reported by \citet{Elias85}, who find that the quantity $[(V - I)_{\rm RSG} +
{\rm BC}_V]$ lies in the remarkably tight range $0.88 \rightarrow 1.0$ for
supergiant stars of spectral types K3 -- M4 (i.e., RSG).  Again, because we
wish to set the most conservative lower detection limit, we assign $(V -
I)_{\rm RSG} + {\rm BC}_V = 0.88$, which is the value obtained for an M4
supergiant star.  Finally, we set $I = 24.4 \pm 0.2$ mag, as derived in
\S~\ref{sec:2.3}.

With these values, we derive $M_{\rm bol} = -6.54 \pm 0.32$ mag as the limiting
bolometric magnitude, above which any RSG would have been detected in our
pre-SN image.  This corresponds to a $3\ \sigma$ detection threshold of $M_{\rm
bol} = -7.50$ mag, which translates to $\log L/L_\odot = 4.90$.  If we adopt a
maximum systematic uncertainty of $0.2$ dex in the theoretical stellar model
endpoints, then the final $3\ \sigma$ lower bound on the luminosity of a RSG
that would have been confidently detected in our pre-SN image is $\log
L/L_\odot = 5.10$.  From Figure~\ref{fig:2} this corresponds to an upper bound
on the progenitor mass of $M_{\rm ZAMS} = 15 {\rm\ M}_\odot$, and we therefore
conclude that any RSG progenitor with an initial mass greater than this value
would have been detected using our analysis procedure.

Applying a similar analysis to the V1 image results in a detection threshold of
$\log L/L_\odot = 6.30$, which unfortunately does not rule out any progenitors
up to $200 {\rm\ M}_\odot$, the highest progenitor mass considered by the STARS
models.  Our most restrictive limit thus comes from the I1 image.

\section{Conclusions}
\label{sec:3}

We analyze two pre-SN and three new post-SN high-resolution images of the site
of the Type II-Plateau supernova SN~2006my in an effort to either detect the
progenitor star or to constrain its properties.  Our primary result is that we
do not detect an isolated stellar object at the location of SN~2006my in either
of the two pre-SN images.  From our image registration, we therefore do not
confirm the association found by L07 between a stellar source (Source 1) in
pre-SN $I$-band images (I1) and the location of SN~2006my.  Using new
high-resolution post-SN images, we derive an offset between SN~2006my and
Source 1 of $\geq 0.08^{\prime\prime}$ from the SN location, which represents a
confidence level of non-association of more than $96\%$ from our most liberal
estimates of the image transformation and measurement uncertainties.  Through
artificial star tests carried out at the precise location of SN~2006my in image
I1, we derive a $3\ \sigma$ upper bound on the mass of the progenitor of
SN~2006my of $M_{\rm ZAMS} = 15 {\rm\ M}_\odot$ from the STARS stellar
evolutionary models.

Although considered unlikely, we can not rule out the possibility that {\it
part} of the light comprising Source 1, which exhibits some extension relative
to other point sources in image I1, or {\it part} of the light contributing to
Source 2, a definitively extended source detected in pre-SN $V$-band images
(V1) that has overlap with the SN~2006my location, may be due to the progenitor
of SN~2006my.  Only additional, high-resolution observations of the site taken
after SN~2006my has faded beyond detection can confirm or reject these
possibilities.

\acknowledgments

We thank an anonymous referee for helpful comments that improved the
manuscript.  A.G. acknowledges support by the Benoziyo Center for Astrophysics,
a research grant from Peter and Patricia Gruber Awards, and the William Z. and
Eda Bess Novick New Scientists Fund at the Weizmann Institute.  A.L.K. is
supported by a NASA Origins grant to L. Hillenbrand.  This research has made
use of the NASA/IPAC Extragalactic Database (NED), which is operated by the Jet
Propulsion Laboratory, California Institute of Technology, under contract with
NASA.  We wish to recognize and acknowledge the very significant cultural role
and reverence that the summit of Mauna Kea has always had within the indigenous
Hawaiian community.  We are most fortunate to have the opportunity to conduct
observations from this mountain.

\clearpage

\appendix
\section{APPENDIX}
\begin{center}
\bf{ON THE CONFIDENCE LEVEL OF NON-ASSOCIATION OF POINT SOURCES}
\end{center}

Determining the significance of positional offsets between objects identified
in two images is of paramount importance when assessing the potential
association between a progenitor star (in a pre-explosion image) and a
supernova (in a post-explosion image), where one image has been transformed
into the pixel frame of the other.  Six sources of uncertainty are typically
identified: Positional uncertainty of the putative progenitor star (in both $x$
and $y$), positional uncertainty of the SN (in both $x$ and $y$), and the
uncertainty in the transformation (also in both $x$ and $y$).  In much of the
past work involving SN progenitors (hereafter, the ``traditional'' approach),
offset significance (i.e., how many ``sigma'' away the two objects are) has
been derived by taking the measured radial offset between the two objects, and
then dividing this value by the quadrature sum of all of the uncertainties.  A
formal analysis, however, finds this approach to be somewhat in error.

Consider the two-dimensional Gaussian that describes the uncertainty in an SN's
measured position ($x, y$) compared with that of a putative progenitor star
($x_0, y_0$):

$$p(x,y)\, dx\, dy\, = \frac{1}{2\pi\sigma_x\sigma_y}e^{-(x -
  x_0)^2/2\sigma_x^2}e^{-(y - y_0)^2/2\sigma_y^2}\,dx\,dy,$$

\noindent where $\sigma_{x(y)}$ represents the quadrature sum of the
measurement and transformation errors in $x(y)$.  Converting this to a
normalized form by defining $\Sigma_x \equiv (x - x_0)/\sigma_x, \Sigma_y
\equiv (y - y_0)/\sigma_y$, yields:

$$p(\Sigma_x,\Sigma_y)\,d\Sigma_x\,d\Sigma_y =
\frac{1}{2\pi\sigma_x\sigma_y}e^{-(\Sigma_x^2/2)}e^{-(\Sigma_y^2/2)}(\sigma_x\,d\Sigma_x)(\sigma_y\,d\Sigma_y),$$

\noindent which becomes the expected:

$$p(\Sigma_x,\Sigma_y)\,d\Sigma_x\,d\Sigma_y =
\frac{1}{2\pi}e^{-(\Sigma_x^2/2)}e^{-(\Sigma_y^2/2)}d\Sigma_x\,d\Sigma_y.$$

\noindent We convert this normalized Gaussian into radial form with a further
change of coordinates ($\Sigma_x \equiv \rho \cos \theta, \Sigma_y \equiv \rho
\sin \theta$) and a little algebra to yield:

$$p(\rho,\theta)\,d\rho\,d\theta = \frac{1}{2\pi}
e^{-\rho^2/2}\rho\,d\rho\,d\theta,$$

\noindent or ignoring the angular part of the distribution,

$$p(\rho)\,d\rho = \rho e^{-\rho^2/2}\,d\rho.$$

\noindent The total integrated probability for $\rho < Q$ (where $Q$ is equal
to, e.g., $[\Sigma_x^2 + \Sigma_y^2]^{1/2}$), then, is

\begin{equation}
p(\rho < Q) = 1 - e^{-(Q^2/2)}.
\label{eqn:a}
\end{equation}

A simple example serves to highlight the differences between calculating offset
significance in the ``traditional'' manner versus calculating it according to
equation~(\ref{eqn:a}).  Suppose a progenitor star candidate and a SN are
measured to be located at pixel locations [10,10] and [11,11] on pre-SN and
(transformed) post-SN images, respectively, with total uncertainties
(quadrature sum of all measurement and transformation uncertainties) of
$\sigma_x = 1$, $\sigma_y = 1$.  In the traditional approach, this would
represent an offset of $1.41 \pm 1.41$ pixels, or a separation significant at
the $1\sigma$ level, which implies a confidence level of non-association of
$68\%$.

However, if we calculate the total integrated probability for $\rho < 1.41$
according to equation~(\ref{eqn:a}) we find $p(\rho < 1.41) = 1 -
\exp[-(1.41^2/2)] = 0.63$, which implies a confidence level of non-association
of $63\%$.  Because we are dealing with positions and uncertainties in two
dimensions, it thus appears most precise to state the result as follows: ``The
two objects are located $1.41\sigma$ away from each other for a two-dimensional
Gaussian, which indicates non-association at the $63\%$ confidence level.''  We
therefore express all of our association significances in this paper in this
manner.


\begin{thebibliography}{29}
\expandafter\ifx\csname natexlab\endcsname\relax\def\natexlab#1{#1}\fi

\bibitem[{{Anderson} \& {King}(2003)}]{Anderson03} 
{Anderson}, J., \& {King}, I.~R.  2003, \pasp, 115, 113 

\bibitem[{{Asplund} {et~al.}(2004)}]{Asplund04} 
{Asplund}, M., {Grevesse}, N., {Sauval}, A.~J., {Allende Prieto}, C., \&  
{Kiselman}, D.  2004, \aap, 417, 751 

\bibitem[{{Bertin} \& {Arnouts}(1996)}]{Bertin96} 
{Bertin}, E., \& {Arnouts}, S.  1996, \aaps, 117, 393 

\bibitem[{{Dolphin}(2000{\natexlab{a}})}]{Dolphin00b} 
{Dolphin}, A.~E.  2000{\natexlab{a}}, \pasp, 112, 1397 

\bibitem[{{Dolphin}(2000{\natexlab{b}})}]{Dolphin00a} 
---. 2000{\natexlab{b}}, \pasp, 112, 1383 

\bibitem[{{Dolphin} \& {Kennicutt}(2002)}]{Dolphin02a} 
{Dolphin}, A.~E., \& {Kennicutt}, Jr. 2002, \aj, 123, 207 

\bibitem[{{Eggleton}(1971)}]{Eggleton71} 
{Eggleton}, P.~P.  1971, \mnras, 151, 351 

\bibitem[{{Eldridge} {et~al.}(2007)}]{Eldridge07} 
{Eldridge}, J.~J., {Mattila}, S., \& {Smartt}, S.~J.  2007, \mnras, 376, 
L52 

\bibitem[{{Eldridge} \& {Tout}(2004)}]{Eldridge04} 
{Eldridge}, J.~J., \& {Tout}, C.~A.  2004, \mnras, 348, 201 

\bibitem[{{Elias} {et~al.}(1985)}]{Elias85} 
{Elias}, J.~H., {Frogel}, J.~A., \& {Humphreys}, R.~M.  1985, \apjs, 57, 91 

\bibitem[{{Filippenko}(1997)}]{Filippenko97} 
{Filippenko}, A.~V.  1997, \araa, 35, 309 

\bibitem[{{Gal-Yam} {et~al.}(2005)}]{Galyam05} 
{Gal-Yam}, A., {et~al.} 2005, \apjl, 630, L29 

\bibitem[{{Heger} \& {Langer}(2000)}]{Heger00} 
{Heger}, A., \& {Langer}, N.  2000, \apj, 544, 1016 

\bibitem[{{Holtzman} {et~al.}(1995)}]{Holtzman95a} 
{Holtzman}, J.~A., {Burrows}, C.~J., {Casertano}, S., {Hester}, J.~J.,  
{Trauger}, J.~T., {Watson}, A.~M., \& {Worthey}, G.  1995, \pasp, 107, 1065 

\bibitem[{{Leonard} {et~al.}(2003)}]{Leonard10} 
{Leonard}, D.~C., {Kanbur}, S.~M., {Ngeow}, C.~C., \& {Tanvir}, N.~R.  
2003,  \apj, 594, 247 

\bibitem[{{Li} {et~al.}(2006)}]{Li06a} 
{Li}, W., {Van Dyk}, S.~D., {Filippenko}, A.~V., {Cuillandre}, J.-C., 
{Jha},  S., {Bloom}, J.~S., {Riess}, A.~G., \& {Livio}, M.  2006, \apj, 
641, 1060 

\bibitem[{{Li} {et~al.}(2007)}]{Li07} 
{Li}, W., {Wang}, X., {Van Dyk}, S.~D., {Cuillandre}, J.-C., {Foley}, 
R.~J., \&  {Filippenko}, A.~V.  2007, \apj, 661, 1013 (L07)

\bibitem[{{Maund} {et~al.}(2005)}]{Maund05} 
{Maund}, J.~R., {Smartt}, S.~J., \& {Danziger}, I.~J.  2005, \mnras, 364, 
L33 

\bibitem[{{Meynet} \& {Maeder}(2000)}]{Meynet00} 
{Meynet}, G., \& {Maeder}, A.  2000, \aap, 361, 101 

\bibitem[{{Nakano} \& {Itagaki}(2006)}]{Nakano06} 
{Nakano}, S., \& {Itagaki}, K.  2006, CBET, 756, 1 

\bibitem[{{Pilyugin} {et~al.}(2004)}]{Pilyugin04} 
{Pilyugin}, L.~S., {V{\'{\i}}lchez}, J.~M., \& {Contini}, T.  2004, \aap, 
425,  849 

\bibitem[{{Schawinski} {et~al.}(2008)}]{Schawinski08} 
{Schawinski}, K., {et~al.} 2008, Science, 321, 223

\bibitem[{{Schechter} {et~al.}(1993)}]{Schechter93} 
{Schechter}, P.~L., {Mateo}, M., \& {Saha}, A.  1993, \pasp, 105, 1342 

\bibitem[{{Smartt} {et~al.}(2008)}]{Smartt08} 
{Smartt}, S.~J., {Eldridge}, J.~J., {Crockett}, R.~M., \& {Maund}, J.~R.  
2008,  \mnras, submitted (ArXiv e-prints 0809.0403)

\bibitem[{{Smartt} {et~al.}(2004)}]{Smartt04} 
{Smartt}, S.~J., {Maund}, J.~R., {Hendry}, M.~A., {Tout}, C.~A., {Gilmore}, 
 G.~F., {Mattila}, S., \& {Benn}, C.~R.  2004, Science, 303, 499 

\bibitem[{{Solanes} {et~al.}(2002)}]{Solanes02} 
{Solanes}, J.~M., {Sanchis}, T., {Salvador-Sol{\'e}}, E., {Giovanelli}, R., 
\&  {Haynes}, M.~P.  2002, \aj, 124, 2440 

\bibitem[{{Stanishev} \& {Nielsen}(2005)}]{Stanishev06} 
{Stanishev}, V., \& {Nielsen}, T.~B.  2006, CBET, 737, 1 

\bibitem[{{Stetson}(1987)}]{Stetson87} 
{Stetson}, P.~B.  1987, \pasp, 99, 191 

\bibitem[{{Van Dyk} {et~al.}(2003)}]{Vandyk03} 
{Van Dyk}, S.~D., {Li}, W., \& {Filippenko}, A.~V.  2003, \pasp, 115, 1289 

\bibitem[{{Wizinowich} {et~al.}(2006)}]{Wizinowich06} 
{Wizinowich}, P.~L., {et~al.} 2006, \pasp, 118, 297 

\end{thebibliography}

\clearpage


\clearpage
\begin{deluxetable}{ccccccccc}
\tabletypesize{\tiny}
\rotate
\tablenum{1}
\tablewidth{525pt}
\tablecaption{Observations of NGC 4651}
\tablehead{
\colhead{ } &
\colhead{ } &
\colhead{ } &
\colhead{ } &
\colhead{Exposure Times}  &
\colhead{ } &
\colhead{Plate Scale} &
\colhead{FWHM } &
\colhead{Combined Image}\\
\colhead{Telescope} &
\colhead{Instrument} &
\colhead{Data Archive Name} &
\colhead{UT Date} &
\colhead{(s)} &
\colhead{Filter } &
\colhead{(arcsec pixel$^{-1}$) } &
\colhead{(arcsec) } &
\colhead{Designation} }

\startdata

HST     & WFPC2       & u2dt0901/2/3t   & 1994 May 20    & $60 + 300 + 300$        & F555W       & 0.10 & 0.15 & V1 \\
HST     & WFPC2       & u2dt0904/5/6t   & 1994 May 20    & $60 + 300 + 300$        & F814W       & 0.10 & 0.15 & I1 \\
Keck II & LGS/NIRC2   & \nodata         & 2006 Nov 28    & $10 \times 6 \times 10$ & ${\rm K}_p$ & 0.04 & 0.10 & K1 \\
HST     & WFPC2       & u9ox0301/2/3/4m & 2007 Apr 26/27 & $4 \times 300$          & F555W       & 0.05 & 0.08 & V2 \\
HST     & WFPC2       & u0ox0305/6m     & 2007 Apr 27    & $500 + 700$             & F814W       & 0.05 & 0.08 & I2 \\

\enddata \tablecomments{\ SN~2006my in NGC~4651 was discovered on 2006 November 8.82
UT \citep{Nakano06}.  Pre-explosion {\it HST}/WFPC2 images (V1 and I1) were
obtained as part of a kinematic study of the core of NGC 4651 (GO 5375; PI:
Rubin).  Post-explosion image K1 was reduced according to the methods detailed
by
\citet{Galyam05}. Details on the reduction and analysis of all images obtained
with {\it HST}/WFPC2 are given in the text.  Plate scale is pixel size of the
detector chip on which the site of SN~2006my is located (i.e., WF2 for V1 and
I1, PC for V2 and I2, and wide-field channel for K1), and FWHM is the measured
full width at half-maximum of a point source in each final, combined image.  We
note that the scale of the wide camera ($ 40^{\prime\prime} \times
40^{\prime\prime}$) is larger than the isoplanatic patch size, which yields a
variable PSF shape over the full image. }

\label{tab:1} 
\end{deluxetable} 
\clearpage 


\clearpage
\begin{deluxetable}{ccccccccc}
\tabletypesize{\small}
\tablenum{2}
\tablewidth{510pt}
\tablecaption{Image Transformations to LGS (Image K1) Pixel Grid}
\tablehead{
\colhead{} &
\colhead{} &
\colhead{} &
\colhead{} &
\multicolumn{2}{c}{Measurement Uncertainty}  &
\colhead{} &
\multicolumn{2}{c}{Transformation Uncertainty}  \\
\cline{5-6} \cline{8-9}
\colhead{Image } &
\colhead{Object } &
\colhead{X } &
\colhead{Y } &
\colhead{$\sigma_x$}  &
\colhead{$\sigma_y$}  &
\colhead{} &
\colhead{$\sigma_x$}  &
\colhead{$\sigma_y$}  }

\startdata

I1 & Source 1  & 579.55 & 543.28 & 0.34 & 0.04 & & \nodata & \nodata \\
K1 & SN 2006my & 581.55 & 542.72 & 0.05 & 0.03 & & 0.30    & 0.27    \\
I2 & SN 2006my & 581.47 & 543.07 & 0.05 & 0.02 & & 0.21    & 0.22    \\
V2 & SN 2006my & 581.69 & 543.07 & 0.02 & 0.02 & & 0.23    & 0.25    \\
V1 & Source 2  & 581.95 & 542.55 & \nodata & \nodata & & 0.15 &
0.15 \\

\enddata \tablecomments{\ Location of sources identified in Fig.~\ref{fig:1}
in images listed in Table~\ref{tab:1}, following transformation of all images
to the K1 pixel grid.  All transformation uncertainties are given in pixel
units (with a plate scale of $0.04^{\prime\prime} {\rm\ pixel}^{-1}$; see
Table~\ref{tab:1}) and represent the rms residuals following registration of
each image to image I1 transformed onto the K1 grid.  Because Source~2 is
extended (and not symmetric), we do not assign a measurement uncertainty on its
position.  Measurement uncertainties were derived as discussed in the text
(\S~\ref{sec:2.2}).  Note the particularly large uncertainty in the {\it x}
coordinate of Source 1 in transformed image I1, likely due to the slight
extension of the object along the {\it x} (i.e., east-west) direction, as
discussed in the text (\S~\ref{sec:2.2}) and evident in Fig.~\ref{fig:1}a.}

\label{tab:2}
\end{deluxetable}

\clearpage



\clearpage
\begin{deluxetable}{ccccccccc}
\tabletypesize{\small}
\tablenum{3}
\tablewidth{510pt}
\tablecaption{Image Transformations to 1994 WFPC2 (Images V1 and I1) Pixel Grid}
\tablehead{
\colhead{} &
\colhead{} &
\colhead{} &
\colhead{} &
\multicolumn{2}{c}{Measurement Uncertainty}  &
\colhead{} &
\multicolumn{2}{c}{Transformation Uncertainty}  \\
\cline{5-6} \cline{8-9}
\colhead{Image } &
\colhead{Object } &
\colhead{X } &
\colhead{Y } &
\colhead{$\sigma_x$}  &
\colhead{$\sigma_y$}  &
\colhead{} &
\colhead{$\sigma_x$}  &
\colhead{$\sigma_y$}  }

\startdata

I1 & Source 1  & 410.23 & 158.59 & 0.29 & 0.29 & & \nodata & \nodata \\
V1 & Source 2  & 410.31 & 159.58 & \nodata & \nodata & & \nodata & \nodata \\
I2 & SN 2006my & 410.18 & 159.41 & 0.03 & 0.03 & & 0.15    & 0.12    \\
V2 & SN 2006my & 410.20 & 159.51 & 0.03 & 0.03 & & 0.12    & 0.09    \\

\enddata \tablecomments{\ Location of sources identified in Fig.~\ref{fig:1}
in images defined in Table 1.  Transformation uncertainties are given in pixel
units (with a plate scale of $0.10^{\prime\prime} {\rm\ pixel}^{-1}$; see
Table~\ref{tab:1}) and represent the rms residuals following registration of I2
onto I1 and V2 onto V1.  The locations of Source 1 and Source 2 in images I1
and V1, respectively, are those reported by {\it hstphot}.  The locations of
SN~2006my in the V1 and I1 pixel grids were derived by taking the {\it
hstphot}-reported coordinates of SN~2006my in images V2 and I2 and transforming
them using the {\it geoxytran} task in IRAF to the V1 and I1 pixel grids,
respectively.  Measurement uncertainties for all object locations are those
suggested by \citet{Dolphin00a}.  Because Source~2 is extended (and not
symmetric), we do not assign a measurement uncertainty on its position.}

\label{tab:3}
\end{deluxetable}

\clearpage


\clearpage

\clearpage

\end{document}